\documentclass[12pt]{article}
\usepackage{qMean}
\begin{document}
\title{Quantum Circuit for Calculating\\
Mean Values\\
Via Grover-like Algorithm}

\author{Robert R. Tucci\\
        P.O. Box 226\\
        Bedford,  MA   01730\\
        tucci@ar-tiste.com}

\date{\today}
\maketitle
\vskip2cm
\section*{Abstract}
In this paper,
we give a quantum circuit for
calculating the mean value
of a function $A(x^n)\in \CC$,
where $x^n\in \{0,1\}^n$.
Known classical algorithms for
calculating the mean value of a structureless function
$A(x^n)$
take $\calo(2^n)$ steps.
Our quantum algorithm
is based on a Grover-like algorithm
and it takes
$\calo(\sqrt{2^n})$ steps.
Our algorithm differs
significantly
from previously proposed quantum
algorithms for
calculating the mean
 value of
a function via Grover's algorithm.

\newpage

\section{Introduction}

In this paper,
we give a quantum circuit for
calculating the mean value
of a function $A(x^n)\in \CC$,
where $x^n\in Bool^n$.
Classical algorithms for
calculating the mean value
of a structureless
function $A(x^n)$
take $\calo(2^n)$ steps.
Our quantum algorithm
is based on the original Grover's algorithm (see Ref.\cite{Gro})
or some variant thereof (such as AFGA, described in Ref.\cite{afga}),
and it takes
$\calo(\sqrt{2^n})$ steps.

Previous papers (see Refs.\cite{Gro,Bra1,Bra2})
have proposed algorithms for
finding the mean value of a function
via Grover's algorithm. Our algorithm
differs significantly from theirs.
One big difference is that
our algorithm encodes the final
answer in the amplitude of a state
whereas theirs encodes it in the
quantum numbers of a state.
A good analogy is to say
that we use something akin to
amplitude modulation (AM radio,
where the signal is encoded in the amplitude)
whereas they use something
akin to frequency modulation (FM
radio, where the signal is encoded
in the frequencies).

This paper assumes that
the reader has already read
most of Ref.\cite{qSym}
by Tucci. Reading that
previous paper is essential
to understanding this one
because
this paper applies
AM techniques (what we call
targeting two hypotheses and blind targeting) described in
that previous paper.
\section{Notation and Preliminaries}
Most of the notation that will be
used in this paper has already been
explained in previous papers by Tucci.
See, in particular, Sec.2
(entitled ``Notation and Preliminaries") of Ref.\cite{qSym}.
\section{Quantum Circuit For
Calculating Mean Values}
In this section,
we will give a quantum
circuit for calculating the
mean value of a
probability amplitude $A(x^n)$
where $x^n\in Bool^n$.
Our algorithm can also
be used to find the
mean value of more general functions
using the method given in Appendix C
 of Ref.\cite{qSym}.

For $x^n\in Bool^n$,
and a {\it normalized} $n$-qubit state $\ket{\psi}$,
define

\beq
\ket{\psi}_{\alpha^{n}}=
\sum_{x^{n}}
A(x^{n})\ket{x^{n}}_{\alpha^{n}}
\;,
\eeq
and

\beq
\ol{A}=
\frac{1}{2^n}\sum_{x^n} A(x^n)
\;.
\eeq
Note that this function $A()$
is not completely general since

\beq
 \sum_{x^{n}}|A(x^n)|^2=1
\;.
\eeq

We will assume that
we know how to compile
$\ket{\psi}_{\alpha^{n}}$
(i.e., that
we can construct it starting
from $\ket{0^n}_{\alpha^{n}}$
using a sequence of
elementary operations.
Elementary operations are
operations that act on a few (usually 1,2 or 3)
qubits at a time,
such as qubit rotations
and CNOTS.)
Multiplexor techniques for doing
such compilations
are discussed in Ref.\cite{tuc-multiplexor}.
If $n$
is very large,
our algorithm will
be useless unless
such a compilation
is of polynomial efficiency,
meaning that
its number of elementary
operations grows as poly($n$).

For concreteness,
we will use $n=3$
henceforth in this section,
but it will be obvious
how to draw
an analogous
circuit
for arbitrary $n$.

\begin{figure}[h]
    \begin{center}
    \epsfig{file=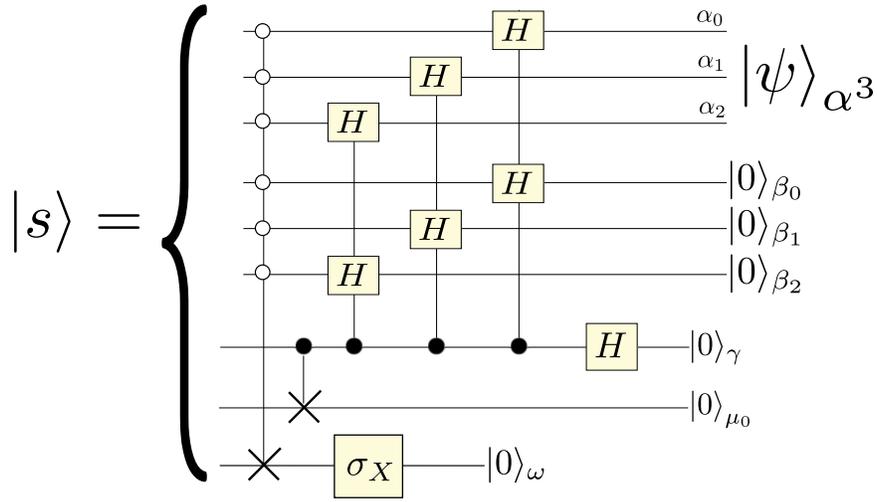, width=4.5in}
    \caption{Circuit for
    generating $\ket{s}$
    used in AFGA to calculate
    mean value of $A(x^3)$.
    }
    \label{fig-qMean-ckt}
    \end{center}
\end{figure}

We want
all
horizontal lines
in Fig.\ref{fig-qMean-ckt}
to represent qubits.
Let $\alpha = \alpha^3$,
$\beta = \beta^3$.

Define

\beq
T(\alpha,\beta)=
\prod_{j=0}^2
\left\{
H(\alpha_j)
H(\beta_j)
\right\}
\;,
\eeq

\beq
\pi(\alpha)=
\prod_{j=0}^2
P_0(\alpha_j)
\;,
\eeq
and

\beq
\pi(\beta)=
\prod_{j=0}^2
P_{0}(\beta_j)
\;.
\eeq

Our method
for
calculating
the mean value of $A(x^3)$
consists of applying the algorithm
AFGA\footnote{As discussed
 in Ref.\cite{qSym},
 we recommend the AFGA
 algorithm, but Grover's original
 algorithm (see Ref.\cite{Gro}) or any other
 Grover-like algorithm
 will also work
 here, as long as it
 drives a
 starting state $\ket{s}$
 to a target state $\ket{t}$.} of Ref.\cite{afga}
in the way that was described in
Ref.\cite{qSym},
using the techniques
of targeting two hypotheses
and blind targeting.
As in Ref.\cite{qSym},
when we apply AFGA in this section,
we will use a sufficient target $\ket{0}_\omega$.
All that remains for
us to do to
fully specify our
circuit for calculating
 the mean value of $A(x^3)$
is to give a circuit for
generating $\ket{s}$.

A circuit for generating
$\ket{s}$ is given by
Fig. \ref{fig-qMean-ckt}.
Fig.\ref{fig-qMean-ckt}
is equivalent to saying that

\beq
\ket{s}_{\mu,\nu,\omega}=
\sigma_X(\omega)^{
\pi(\beta)
\pi(\alpha)}
\frac{1}{\sqrt{2}}
\left[
\begin{array}{l}
T(\alpha,\beta)
\begin{array}{l}
\ket{\psi}_{\alpha}
\\
\ket{0^3}_\beta
\end{array}
\\
\ket{1}_\gamma
\\
\ket{1}_{\mu_0}
\\
\ket{1}_\omega
\end{array}
+
\begin{array}{l}
\ket{\psi}_{\alpha}
\\
\ket{0^3}_\beta
\\
\ket{0}_\gamma
\\
\ket{0}_{\mu_0}
\\
\ket{1}_\omega
\end{array}
\right]
\;.
\eeq

\begin{claim}

\beq
\ket{s}_{\mu,\nu,\omega}=
\begin{array}{c}
z_1 \ket{\psi_1}_{\mu}
\\
\ket{1}_{\nu}
\\
\ket{0}_\omega
\end{array}
+
\begin{array}{c}
z_0 \ket{\psi_0}_{\mu}
\\
\ket{0}_{\nu}
\\
\ket{0}_\omega
\end{array}
+
\begin{array}{c}
\ket{\chi}_{\mu,\nu}
\\
\ket{1}_\omega
\end{array}
\;,
\eeq
for some unnormalized state
$\ket{\chi}_{\mu,\nu}$,
where

\beq
\begin{array}{|c|c|}
\hline
\ket{\psi_1}_{\mu}=
\begin{array}{l}
\ket{0^3}_{\alpha}
\\
\ket{1}_{\mu_0}
\end{array}
&
\ket{\psi_0}_{\mu}=
\begin{array}{l}
\ket{0^3}_{\alpha}
\\
\ket{0}_{\mu_0}
\end{array}
\\
\ket{1}_{\nu}=
\left[
\begin{array}{r}
\ket{0^3}_{\beta}
\\
\ket{1}_{\gamma}
\end{array}
\right]
&
\ket{0}_{\nu}=
\left[
\begin{array}{r}
\ket{0^3}_{\beta}
\\
\ket{0}_{\gamma}
\end{array}
\right]
\\
\hline
\end{array}
\;,
\eeq

\beq
z_1= \frac{1}{\sqrt{2}}\left[\frac{1}{2^3}
\sum_{x^3}\av{x^3|\psi}\right]
\;,
\eeq

\beq
z_0 =
\frac{1}{\sqrt{2}}
\av{0^3|\psi}
\;,
\eeq

\beq
\frac{|z_1|}{|z_0|} = \sqrt{\frac{P(1)}{P(0)}}
\;.
\label{eq-z1-z0-again}
\eeq
\end{claim}
\proof

Recall that for any
quantum systems $\alpha$ and $\beta$,
any
unitary operator $U(\beta)$
and any
projection operator $\pi(\alpha)$,
one has

\beq
U(\beta)^{\pi(\alpha)}=
(1-\pi(\alpha)) + U(\beta)\pi(\alpha)
\;.
\label{eq-u-pi-id}
\eeq
Applying identity Eq.(\ref{eq-u-pi-id}) with $U=\sigma_X(\omega)$
yields:

\beqa
\ket{s} &=&
\sigma_X(\omega)^{\pi(\beta)\pi(\alpha)}\ket{s'}
\\
&=&
\sigma_X(\omega)\pi(\beta)\pi(\alpha)\ket{s'}
+
\begin{array}{l}
\ket{\chi}_{\mu,\nu}
\\
\ket{1}_\omega
\end{array}
\\
&=&
\frac{1}{\sqrt{2}}
\left[
\begin{array}{l}
\pi(\beta)
\pi(\alpha)T(\alpha,\beta)
\begin{array}{l}
\ket{\psi}_{\alpha}
\\
\ket{0^3}_\beta
\end{array}
\\
\ket{1}_\gamma
\\
\ket{1}_{\mu_0}
\\
\ket{0}_\omega
\end{array}
+
\begin{array}{l}
A(0^3)\ket{0^3}_{\alpha}
\\
\ket{0^3}_\beta
\\
\ket{0}_\gamma
\\
\ket{0}_{\mu_0}
\\
\ket{0}_\omega
\end{array}
\right]
+
\begin{array}{l}
\ket{\chi}_{\mu,\nu}
\\
\ket{1}_\omega
\end{array}
\;.
\eeqa
Applying identity Eq.(\ref{eq-u-pi-id}) with $U=\sigma_X(\beta_j)$
yields:

\beq
\pi(\beta)\pi(\alpha)
T(\alpha,\beta)
\begin{array}{l}
\ket{\psi}_{\alpha}
\\
\ket{0^3}_\beta
\end{array}
=
\begin{array}{l}
\ket{0^3}_\alpha \frac{1}{2^3}\sum_{x^3} A(x^3)
\\
\ket{0^3}_\beta
\end{array}
\;.
\eeq
\qed

Note that the amplitude $z_0$
 for the null hypothesis is $\ket{\psi}$
dependent. This is contrary to the
Mobius Transform algorithm of Ref.\cite{qMob} where the $z_0$ is
$\ket{\psi^-}$ independent.

Note also that
for this method of finding the mean value
of $A(x^n)$ to work well,
$\ol{A}$ and $A(0^n)$ must be of
``comparable" size, for if $|z_0|^2<<|z_1|^2$, then, by Eq.(\ref{eq-z1-z0-again}),
$P(0)<<P(1)$, and it will take
an unreasonable amount of time
to get a bunch of null events.
Of course, the circuit for generating
$\ket{s}$ can be changed easily
so that $z_0$ is proportional to
$A(y^n)$ instead of $A(0^n)$,
where $y^n$ is any element of $Bool^n$.


\begin{thebibliography}{99}

\bibitem{Gro}Lov K. Grover,
``Quantum computers can search
rapidly by using almost any transformation",
arXiv:quant-ph/9712011

\bibitem{afga}
R.R. Tucci, ``An Adaptive, Fixed-Point Version of Grover's Algorithm", arXiv:1001.5200

\bibitem{Bra1}
G. Brassard, P. Hoyer, M. Mosca, and A. Tapp, ``Quantum amplitude amplification and estimation",  arXiv:quant-ph/0005055

\bibitem{Bra2}
G. Brassard, F. Dupuis, S. Gambs, and A. Tapp, ``An optimal quantum algorithm to approximate the mean and its application for approximating the median of a set of points over an arbitrary distance",  arXiv:1106.4267

\bibitem{qSym}
R.R. Tucci,
``Quantum Circuit for Calculating
Symmetrized Functions
Via Grover-like Algorithm",
arXiv:1403.6707

\bibitem{tuc-multiplexor}
R.R. Tucci, ``Code Generator for Quantum Simulated Annealing", arXiv:0908.1633

\bibitem{qMob} R.R. Tucci,
``Quantum Circuit for Calculating
Mobius-like Transforms Via Grover-like Algorithm", arXiv:1403.6910

\end{thebibliography}
\end{document}